\documentclass[12pt]{article}
\usepackage{amsfonts}
\usepackage{amsmath}
\usepackage{amssymb}
\usepackage{graphicx}
\usepackage{color}
\usepackage[all, knot]{xy}
\usepackage{tikz}
\usepackage{array}
\usepackage{hyperref}
\usepackage{ulem}

\usepackage[utf8]{inputenc}
\usepackage{epstopdf}
\usepackage[footnotesize]{caption}
\usepackage{subcaption}
\usepackage{amsthm}
\usepackage{enumitem}
\usepackage{mathrsfs}
\usepackage{mathtools}

\usepackage[margin=3cm]{geometry}

\def \be {\begin{equation}}
\def \ee {\end{equation}}
\def \bea {\begin{eqnarray}}
\def \eea {\end{eqnarray}}
\def \nn {\nonumber}

\def \rr {\raise.35ex\hbox{\small $\prime$}\kern-.17em{\mbox{\large $\imath$}}}

\def \dels {\partial\kern-.6em /\kern.1em}
\def \As {{A\kern-.5em / \kern.5em}}
\def \Ds {D\kern-.7em / \kern.5em}

\def \ks {k\kern-.5em /}
\def \ls {l\kern-.5em /}

\newcommand{\ci}[1]{}

\newcommand{\ba}{\begin{eqnarray}}
\newcommand{\ea}{\end{eqnarray}}
\newcommand{\bal}{\begin{align}}
\newcommand{\eal}{\end{align}}
\newcommand{\bay}[1]{\left(\begin{array}{#1}}
\newcommand{\eay}{\end{array}\right)}

\newcommand{\hide}[1]{}

\newlist{axioms}{enumerate}{2}
\setlist[axioms,1]{label=\textbf{A\arabic{axiomsi}.}, ref=A\arabic{axiomsi}}
\setlist[axioms,2]{label=\textbf{A\arabic{axiomsi}\rlap{\myEnumCounter{axiomsii}}.},%
                   ref=A\arabic{axiomsi}\myEnumCounter{axiomsii},%
                   align=parleft,%
                   leftmargin=0em,%
                   itemsep=1.4ex,%
                   before={\stepcounter{axiomsi}}}

  \usetikzlibrary{decorations.markings}

\begin{document}

\begin{titlepage}
\begin{center}

\textbf{\LARGE
Chaotic-Integrable Transition for Disordered Orbital Hatsugai-Kohmoto Model
\vskip.3cm
}
\vskip .5in
{\large
Ying-Lin Li$^{a}$ \footnote{e-mail address: s1012424@gmail.com},
Chen-Te Ma$^{b}$ \footnote{e-mail address: yefgst@gmail.com}, and 
Po-Yao Chang$^{a}$ \footnote{e-mail address: pychang@phys.nthu.edu.tw}
\\
\vskip 1mm
}
{\sl
$^a$
Department of Physics, National Tsing Hua University, Hsinchu 30013, Taiwan. 
\\
$^b$
Department of Physics and Astronomy, Iowa State University, Ames, Iowa 50011, US.
}
\\
\vskip 1mm
\vspace{40pt}
\end{center}
\begin{abstract}
We have drawn connections between the Sachdev-Ye-Kitaev model and the multi-orbit Hatsugai-Kohmoto model, emphasizing their similarities and differences regarding chaotic behaviors.
The features of the spectral form factor, such as the dip-ramp-plateau structure and the adjacent gap ratio, indicate chaos in the disordered orbital Hatsugai-Kohmoto model. 
One significant conclusion is that the plateau value of the out-of-time-order correlator, whether in the Hatsugai-Kohmoto model, Sachdev-Ye-Kitaev model with two- or four-body interactions, or a disorder-free Sachdev-Ye-Kitaev model, does not effectively differentiate between integrable and chaotic phases in many-body systems.
This observation suggests a limitation in using out-of-time-order correlator plateau values as a diagnostic tool for chaos.
Our exploration of these ideas provides a deeper understanding of how chaos arises in non-Fermi liquid systems and the tools we use to study it.
It opens the door to further questions, particularly about whether there are more effective ways to distinguish between chaotic and integrable phases in these complex systems.
\end{abstract}
\end{titlepage}

\section{Introduction}
\label{sec:1}
\noindent
Non-Fermi liquids (NFLs) are characterized by the absence of well-defined quasiparticles, which distinguishes them from Fermi liquids.
The notable NFL models, such as the {\it Sachdev-Ye-Kitaev} (SYK) model \cite{kitaev2015simple,Polchinski:2016xgd} and the multi-channel Kondo impurity (MCKI) model, both of which exhibit chaotic behavior \cite{Polchinski:2016xgd,Garcia-Garcia:2016mno,Cotler:2016fpe,Larzul:2024qrr}. Recently, renewed interest intertwining of topology and correlations has emerged in a class of exactly solvable many-body fermionic models
with infinite-range interactions~\cite{Manning-Coe:2023etj}, initially proposed by Hatsugai and Komohto (HK) \cite{hatsugai1992exactly}.
These models are described by topologically nontrivial band topology together with the HK interactions, where
topological Mott insulating phases emerge~\cite{Mai:2022mpt, Manning-Coe:2023etj}.
It has been found that the ground states of the HK model do not exhibit the characteristics of Fermi liquids. 
By introducing an energy cost $U$ for double occupancy in each momentum state, the HK model can be expressed as
\begin{align}
\label{eq:HK}
H_{\mathrm{HK}} = \sum_{k\sigma}{ \varepsilon_{k}n_{k\sigma}} + U\sum_{k}{n_{k\uparrow}n_{k\downarrow}}.
\end{align}
Here $\varepsilon_{k}$ denotes the dispersion in a tight binding model with no orbital degrees of freedom, with the band index $k$ in the one-dimensional Brillouin zone (BZ), and $\sigma=\uparrow,\downarrow$ indicates the $z$-component of the electron spin.
The number operator is defined as
$n_{\alpha,\sigma}=c^{\dagger}_{\alpha,\sigma}c_{\alpha,\sigma}$
under the many-body basis.
This paper considers HK-type interactions that impose an energy penalty on doubly-occupied orbital degrees of freedom $\alpha$.
An extension of the HK model, which includes an orbital term \cite{Manning-Coe:2023etj}
\begin{align}
\label{eq:orbit-HK}
H_{\mathrm{OHK}} = & \sum_{k,\alpha<\alpha',\sigma}{ \big(t_{\alpha\alpha'\sigma}(k)c^{\dagger}_{k\alpha\sigma}c_{k\alpha'\sigma}} + \mathrm{H. C.}\big) \notag\\
&- \mu\sum_{k,\alpha}{n_{k\alpha\sigma}} + \sum_{k,\alpha,\alpha'}{U_{\alpha\alpha'}(k)n_{k\alpha\uparrow}n_{k\alpha'\downarrow}},
\end{align}
where we assume
$U_{\alpha,\alpha^{\prime}}(k) =U_{k,\alpha}\delta_{\alpha,\alpha^{\prime}}$
respecting the Pauli principle.
The chemical potential is set to zero for simplicity, i.e., $\mu=0$.
We adjust the permissible orbital number in the half-filling scenario with zero total spin.
Extending the HK model to include orbital degrees of freedom preserves its non-Fermi liquid behavior.
The solvability of the HK model arises from its locality in momentum space.
However, when viewed in position space, the interaction term becomes highly non-local, coupling {\it position-space} points through a four-point function.
This is similar to the SYK model with a four-body interaction (SYK$_{4}$ model). Exploring the chaotic phenomena of the orbital HK model is significant.
\\

\noindent
In classical mechanics, chaotic systems are highly {\it sensitive} to initial conditions, a phenomenon captured by the exponential divergence of nearby trajectories.
However, due to the uncertainty principle, quantum systems governed by wavefunctions and probabilities do not allow for the same kind of trajectory-based analysis.
The difficulty arises because, taking the limit where $\hbar\rightarrow 0$ and the system's non-integrability parameter (which measures how far the system is from being integrable) becomes small, do {\it not} commute \cite{Berry:1977zz}.
This lack of commutativity means that simply shrinking $\hbar$ does not smoothly lead to the classical chaotic behavior one might expect.
This discrepancy complicates efforts to explain classical chaos in a quantum framework.
While quantum systems {\it cannot} exhibit classical chaos in the traditional sense (because their evolution is deterministic via the Schrödinger equation), they do exhibit quantum signatures of chaos, such as:
\begin{itemize}
\item{Energy Level Statistics: In chaotic quantum systems, the distribution of energy levels tends to follow the predictions of random matrix theory (RMT) \cite{Dyson:1962es}, in contrast to the regular, predictable spacing seen in integrable systems \cite{Garcia-Garcia:2017bkg}.}
\item{Wavefunction Behavior: Chaotic quantum systems can display complicated, highly irregular wavefunctions, contrasting with the regular patterns seen in non-chaotic systems \cite{Berry:1977wpp}.}
\item{Quantum Scars: In some cases, wavefunctions of quantum chaotic systems can show localized structures along classical periodic orbits, known as "quantum scars".
}
\end{itemize}
One popular approach to studying quantum chaos involves semi-classical methods, which attempt to bridge the gap between quantum and classical descriptions \cite{Muller:2004nb}.
Gutzwiller's trace formula is an example of relating quantum energy levels to classical periodic orbits \cite{Muller:2004nb}.
\\

\noindent
In the context of quantum chaos, RMT provides a statistical framework to understand the energy level distribution in quantum systems, especially those that exhibit chaotic behavior \cite{Bohigas:1983er}.
In integrable quantum systems (those that have regular, predictable behavior), energy levels tend to cluster in a Poissonian distribution, meaning the gaps between levels are random and independent \cite{Bohigas:1983er}.
However, in chaotic quantum systems, the energy levels tend to avoid each other, leading to {\it level repulsion} \cite{Bohigas:1983er}.
The distribution of energy level spacings in such systems follows the Wigner-Dyson distribution (from RMT) \cite{Atas:2013gvn}.
This describes the probability of finding two nearby energy levels at a certain distance apart.
The distribution is typical of random matrices that are either:
\begin{itemize}
\item{Orthogonal Ensemble (if the system respects time-reversal symmetry);}
\item{Unitary Ensemble (if the system breaks time-reversal symmetry);}
\item{Symplectic Ensemble (for systems with additional symmetries like spin-orbit coupling).}
\end{itemize}
For example, for the Gaussian orthogonal ensemble (GOE), the level spacing follows
$P(s)=\frac{\pi s}{2}\exp\bigg(-\frac{\pi s^2}{4}\bigg)$,
where $s$ is the normalized spacing between neighboring energy levels.
This distribution shows a low probability of two energy levels being very close together or far apart, with a peak indicating some typical spacing between levels.
\\

\noindent
The distribution of energy levels provides the uniquely chaotic properties in a many-body quantum system.
Currently, we lack a general method to study quantum chaos in a few-body system \cite{Kirkby:2021gqc}.
Hence, diagnosing quantum chaos is still challenging.
Another diagnostic method from the energy spectrum is the spectral form factor (SFF), which is the square of the Fourier transform of the empirical spectral density \cite{Brezin:1997rze,Dyer:2016pou}.
With the disorder average, the SFF ($g$), the disconnected part ($g_d$) and the connected part ($g_c$) are defined:
\begin{align}
g(t,\beta) &\equiv \frac{\langle Z(\beta,t)Z^{*}(\beta,t)\rangle_{t_{h},U}}{\langle Z(\beta)\rangle_{t_{h},U}^{2}};\\
g_{d}(t, \beta) &\equiv \frac{\langle Z(\beta,t)\rangle_{t_{h},U}\cdot\langle Z^{*}(\beta,t)\rangle_{t_{h},U}}{\langle Z(\beta)\rangle_{t_{h},U}^{2}};\\
g_{c}(t, \beta) &\equiv g(t, \beta)-g_{d}(t, \beta)
\end{align}
in terms of the partition function, which is the trace of the time-evolved and thermally weighted Hamiltonian \cite{Okuyama:2018yep}
\bea
Z(\beta,t)\equiv\mathrm{Tr}(e^{-\beta H-iHt}), 
\eea
where the inverse temperature is $\beta$.
The disorder average $\langle\cdot\rangle_{t_{h}, U}$ ensures that the SFF is averaged over an ensemble of random couplings ($t_h$ and $U$).
\\

\noindent
In chaotic quantum systems, the SFF is expected to display the characteristic "dip-ramp-plateau" structure \cite{Brezin:1997rze,Dyer:2016pou,Okuyama:2018yep}.
Initially, the SFF decreases (dip) due to interference between energy levels.
As the system evolves, the SFF ramps up, reflecting the emergence of a correlation between energy levels.
Eventually, it reaches a plateau, indicating the saturation of correlations, a feature of quantum chaos.
The disorder parameter or randomness can generate the dip-ramp-plateau to the shape of the SFF in an integrable system \cite{Lau:2018kpa,Lau:2020qnl}.
Therefore, the SFF's shape alone cannot uniquely diagnose quantum chaos \cite{Lau:2018kpa,Lau:2020qnl}.
This limitation suggests that other spectral observables or diagnostics are necessary to capture the chaotic dynamics of few-body systems fully.
\\

\noindent
We discuss the role of sensitivity to initial conditions and out-of-time-order correlations (OTOCs) in distinguishing chaotic and integrable systems, particularly in quantum many-body systems.
While sensitivity to initial conditions is often associated with chaos (as in classical chaos theory), it alone is {\it insufficient} to differentiate chaotic systems from integrable ones.
The system's dynamics can still exhibit sensitivity to initial conditions in integrable systems.
However, they do {\it not} exhibit the irregular behavior typical of chaos \cite{Xu:2019lhc}.
OTOCs are
\bea
C(t) = \langle\lbrack V(0),W(t)\rbrack \lbrack W^{\dagger}(t),V^{\dagger}(0)\rbrack\rangle\geq 0,
\eea
where $W(t)$ and $V(t)$ are operators in the Heisenberg representation, and the angle bracket denotes the thermal average,
a diagnostic tool for detecting the chaotic features in quantum systems.
These correlation functions reflect how operators evolve and fail to commute over time.
In the context of chaos, OTOCs show exponential growth, a hallmark of chaotic behavior.
In the semi-classical limit, OTOCs exhibit exponential growth up to a time scale known as the Ehrenfest time.
The growth rate is characterized by the Lyapunov exponent $\lambda_L$ (the exponent of the OTOCs when it is non-negative), which measures how rapidly initial perturbations grow over time.
In some cases, such as the SYK model, the Lyapunov exponent reaches a theoretical upper bound \cite{Polchinski:2016xgd,Maldacena:2015waa}.
The disorder-free SYK model \cite{Lau:2020qnl}, though, does {\it not} have random matrix behavior, still shows chaotic characteristics with a {\it non-zero} Lyapunov exponent when the system size is large enough \cite{Ozaki:2024wpj}.
Introducing disorder into systems can induce chaotic behavior.
However, it is {\it essential} to differentiate this induced randomness from inherent quantum chaos \cite{Lau:2018kpa,Lau:2020qnl}.
The presence of randomness does {\it not} necessarily mean that the system is chaotic.
In general, exchanging the logarithm and the thermal average influences the Lyapunov exponent \cite{Trunin:2023rwm}.
Therefore, the Lyapunov exponent generally depends on regularization, {\it unless} considering a many-body system.
The late-time behavior of OTOCs, precisely the plateau value of the correlation function, is a topic of ongoing investigation \cite{Markovic:2022jta}.
It could serve as an indicator of chaos in general systems.
However, questions remain about whether this plateau is merely a finite-size artifact or holds deeper physical meaning.
\\

\noindent
This paper examines chaotic behaviors in non-Fermi liquid systems, specifically focusing on the disordered orbital HK model.
By introducing disorder to the model as that $t_{\alpha\alpha^{\prime}\sigma}(k)$ ($t_h$) and $U_{k, \alpha}$ ($U$) follow the Gaussian distribution with the zero mean and the variance of $t_h$ and $U$ can be tuned and studying statistical distributions, the paper provides insights into how chaos emerges.
After the Fourier transformation to the position space, the disordered orbital HK model is similar to the SYK$_2$+SYK$_4$ model.
If the range of the band index goes to infinity, the summation of the momentum index vanishes in the HK model \cite{Mai_2024}.
Hence, this model is simpler.
Key points include using the adjacent gap ratio and spectral form factor to characterize chaos and transition between integrable and chaotic phases.
The paper also calculates OTOCs to explore the impact of regularization on late-time behavior by comparing the disordered orbital HK model to the SYK$_2$, SYK$_4$, and disordered-free SYK models.
Our paper elucidates quantum chaotic behaviors in disordered non-Fermi liquids, particularly through the novel application of statistical tools like the adjacent gap ratio, the SFF, and OTOCs to probe the late-time scale.
We summarize our results as follows:
\begin{itemize}
\item
A phase diagram for the variances of the disorder parameters, $t_h$ and $U$, is constructed using the adjacent gap ratio in the disordered orbital HK model.
This model has the GOE by turning on the four-point term.
The SFF is consistent with chaos in the disordered orbital HK model.
\item The plateau value of the OTOCs varies depending on regularization, indicating that the late-time behavior of the system is sensitive to how regularization is handled.
Unlike the early-time regime, the temperature dependence at late times does not differentiate between integrable and chaotic behavior.
\end{itemize}

\noindent
The organization of this paper is as follows: In Sec.~\ref{sec:2}, we introduce the disordered orbital HK model.
In Sec.~\ref{sec:3}, we study the spectral statistics by calculating the adjacent gap ratio and the spectral form factor.
Sec.~\ref{sec:4} explores the late-time scale of the OTOCs.
Finally, we present our concluding remarks in Sec.~\ref{sec:5}.

\section{Disordered Orbital HK Model}
\label{sec:2}
\noindent
The Hamiltonian of the disordered orbital HK model, after the Fourier transformation
\bea
c^\dagger_{x_1\alpha\sigma} = \sum_{k\in \mathrm{BZ}}{e^{\mathrm{i} k x_1}c^\dagger_{k\alpha\sigma}},
\eea
is given by
\bea
&&
H_{\mathrm{OHK}}
\nn\\
&=& \sum_{x_{1,2},\alpha<\alpha',\sigma}{\sum_{k}{t_{k\alpha\alpha'\sigma}e^{ik(x_{1}-x_{2})}}c^{\dagger}_{x_{1}\alpha\sigma}c_{x_{2}\alpha'\sigma}} + \mathrm{H. C.}
\nn\\
&&
+ \sum_{x_{1,2,3,4}\alpha}\sum_{k}U_{k\alpha}e^{ik(x_{1}-x_{2}+x_{3}-x_{4})}   \notag\\
&& \times c^{\dagger}_{x_{1}\alpha\uparrow} c_{x_{2}\alpha\uparrow}c^{\dagger}_{x_{3}\alpha\downarrow}c_{x_{4}\alpha\downarrow},
\label{eq:orbit-HKrealspace}
\eea
where $x_{1}, x_{2}, x_{3}, x_{4}$ label position space lattice point.
Therefore,
\bea
\sum_{k}{t_{k\alpha\alpha'\sigma}e^{ik(x_{1}-x_{2})}}; \ \sum_{k}{U_{k\alpha}e^{ik(x_{1}-x_{2}+x_{3}-x_{4})}}
\eea
play the role of random variables for two-point and four-point functions in a real space, respectively.
This model is analogous to the SYK$_2$+SYK$_4$ model, described by the Hamiltonian
\begin{align}
H_{\mathrm{SYK2+SYK4}} = &\frac{i}{2!}\sum^{N}_{i_1,i_2=1}{\kappa_{i_1i_2}\chi_{i_1}\chi_{i_2}}  \notag\\
&+ \frac{1}{4!}\sum^{N}_{i_1,i_2,i_3,i_4=1}{J_{i_1i_2i_3i_4}\chi_{i_1}\chi_{i_2}\chi_{i_3}\chi_{i_4}},
\label{Eq:SYK24}
\end{align}
where $\chi_{i_1}$ represents Majorana fermions, and the random couplings $J_{ijkl}$ and $\kappa_{ij}$ follow the Gaussian distribution with the zero mean and the standard deviations $\sqrt{6}J/N^{3/2}$ and $\kappa/\sqrt{N}$, respectively.
We can generally set the fixed-coupling constant $J$ as one because the results only depend on the ratio of two fixed coupling constants $J$ and $\kappa$.
We define $N$ as the number of Majorana fermion fields.
The two-body interaction term is called the SYK$_2$ model, and another four-body interaction is called the SYK$_4$ model.
A chaotic-integrable transition has been observed in the SYK$_2$+SYK$_4$ model \cite{Garcia-Garcia:2017bkg}.
This raises the question of whether a similar transition might occur in the disordered orbital HK model.
Comparing the SYK$_2$ + SYK$_4$ [Eq.~(\ref{Eq:SYK24})] and the orbital HK model [Eq.~(\ref{eq:orbit-HKrealspace})],
the former conserves fermionic parity, while the latter conserves charge.
We expect that a charge-conserving variant of the SYK model, in which  Majorana fermions are replaced by complex fermions, should yield qualitatively similar results to the orbital HK model.
\\

\noindent
The disorder-free SYK model \cite{Lau:2020qnl} is the SYK$_4$ model with the uniform distribution \cite{Ozaki:2024wpj}
\bea
J_{i_1i_2i_3i_4}=1.
\eea
The operators of the OTOCs in the SYK$_2$, SYK$_4$, and disorder-free SYK models are the Majorana fermions.
While in the disorder orbital HK model, the operator of the OTOC is the number operator.
\\

\noindent
Before proceeding with our analysis, we first summarize the known and unknown results, and our results in the non-Fermi liquid models and the related models in Tab. \ref{table:conclusion}.
The "$\beta$ ind." means that the behavior of late-time OTOCs is temperature independent.
For the SYK$_{2}$ and SYK$_4$ models, the analyses are consistent across studies \cite{Cotler:2016fpe,Garcia-Garcia:2017bkg, Lau:2020qnl}.
However, there is a mismatch in the disorder-free SYK$_2$ model between spectral statistics and OTOC analysis \cite{Ozaki:2024wpj}.
This indicated that even in the absence of random matrix theory properties, the system can still exhibit exponential growth in the early-time behavior of OTOCs \cite{Ozaki:2024wpj}.
Besides, for another non-Fermi liquid model, the multi-channel Kodon model, exponential growth in early-time OTOCs was observed in the large-$N$ limit \cite{Larzul:2024qrr}, but its spectral statistics remain unknown.
For our disordered orbital HK model, chaotic behavior is evident from the adjacent gap ratio and SFF.
Lastly, since the late-time OTOCs of SYK$_2$, SYK$_4$, and disorder-free SYK$_2$ are temperature-independent, we argue that the temperature dependence observed in late-time behavior arises from the finite size effect and is not a reliable indicator for many-body chaos \cite{Markovic:2022jta}.
\begin{table*}[ht!]
\centering
\begin{tabular}{|m{2.1cm}||m{2.1cm}|m{3.8cm}|m{2.1cm}|m{2.5cm}|}
\hline
\hspace{1em}Model & Adjacent Gap Ratio & \hspace{1.5cm}SFF & Early-Time OTOCs & Late-Time OTOCs \\
\hline
\hline
\hspace{1em}SYK$_{2}$ & Poisson \cite{Garcia-Garcia:2017bkg} & no linear ramp \cite{Lau:2020qnl} & constant in time \cite{Garcia-Garcia:2017bkg} & \hspace{2em}$\beta$ ind. \\
\hline
\hspace{1em}SYK$_{4}$ & all random matrix \cite{Cotler:2016fpe} & dip-ramp-plateau \cite{Cotler:2016fpe} & exponential growth \cite{Garcia-Garcia:2017bkg} & \hspace{2em}$\beta$ ind. \\
\hline
Disorder-Free SYK & Poisson \cite{Ozaki:2024wpj} & no linear ramp \cite{Ozaki:2024wpj} & exponential growth \cite{Ozaki:2024wpj} & \hspace{2em}$\beta$ ind. \\
\hline
\hspace{1em}MCKI & \hspace{2em}? &  \hspace{4em}? & exponential growth \cite{Larzul:2024qrr} & \hspace{1.2cm}? \\
\hline
Disordered Orbital HK & GOE and Poisson & dip-ramp-plateau (GOE) and no linear ramp (Poisson) & \hspace{2em}? & regularization dep. \\
\hline
\end{tabular}
\caption{Chaotic properties of various non-Fermi liquid models and related models.}
\label{table:conclusion}
\end{table*}

\section{Spectral Statistics}
\label{sec:3}
\noindent
We study the spectral statistics of the disordered orbital HK model by calculating the average adjacent gap ratio \cite{Atas:2013gvn} through the exact diagonalization.
The phase diagrams for the variances of the $t_h$ and $U$ show the GOE and Poisson regimes.
The transition between chaotic and integrable behavior is observed.
The SFF \cite{Brezin:1997rze} demonstrates the dip-ramp-plateau behavior in the GOE regime, which supports our observation.

\subsection{Adjacent Gap Ratio}
\begin{center}
\begin{figure}
\centering
\includegraphics[width=0.95\linewidth]{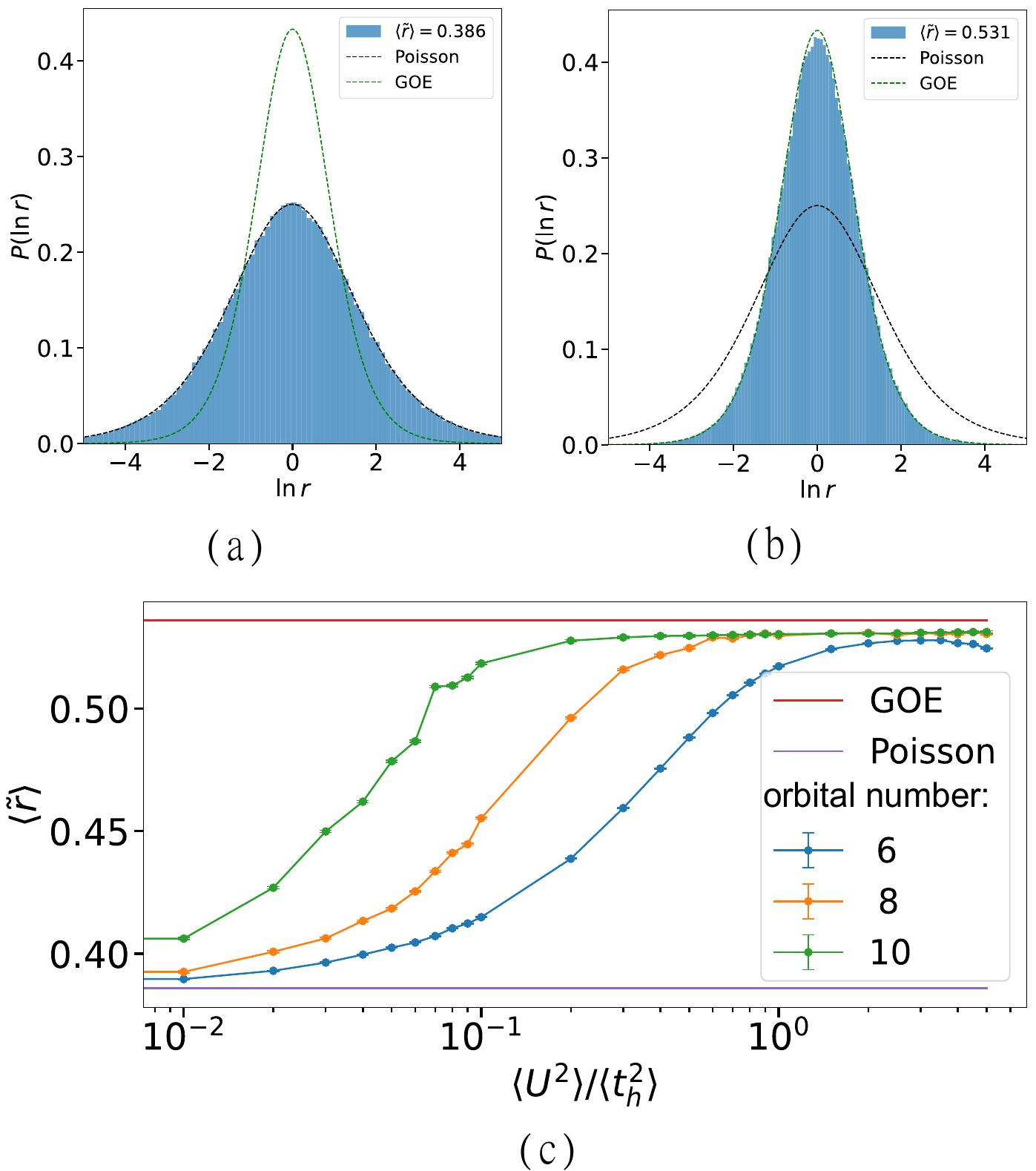}
\caption{Logarithmic ratio distribution of $\ln r$ over 5 random samples for the disordered orbital HK model with 10 orbitals with half-filling and 0 total spin.
(a) Chaotic phase with $\langle \tilde{r} \rangle = 0.531 $, $\langle t_h^2 \rangle = 1 $, and $\langle U^2 \rangle = 3.3$.
(b) Integrable phase with $\langle \tilde{r} \rangle = 0.386$, $\langle t_h^2 \rangle = 1$, and $\langle U^2 \rangle = 0$.
(c) The phase diagram of the average adjacent gap ratio as a variance of $U$ over a variance of $t_h$. 6 orbitals with 5000 random samples; 8 orbitals with 100 random samples; 10 orbitals with 5 random samples.}
\label{fig:level_statistic_N_compare}
\end{figure}
\end{center}

\noindent
To analyze the level statistics, we first collect the eigenenergies $E_{n}$ and arrange them in ascending order
\bea
E_{1}<E_{2}<\cdots.
\eea
We define the level spacing as
\bea
\Delta E_{n} = E_{n+1}-E_{n},
\eea
and the ratios of adjacent level spacings are given by
\bea
r_{n} =\frac{ \Delta E_{n}}{\Delta E_{n+1}}.
\eea
In a generic integrable system, the energy levels are uncorrelated such that the probability distribution of $r_{n}$ follows the Poisson distribution,
\bea
p_{p}(r_n) = \frac{1}{(1+r_n)^{2}}.
\eea
For a quantum chaotic system, the energy levels repel each other, resulting in a probability distribution that follows Wigner-Dyson (WD) statistics for $3\times3$ matrices \cite{Muller:2004nb,Bohigas:1983er},
\bea
p_{W}(r_n) = \frac{1}{A_{\nu}}\frac{(r_n+r_n^{2})^{\nu}}{(1+r_n+r_n^{2})^{1+1.5\nu}},
\eea
where $\nu$ and $A_{\nu}$ are constants depending on the ensemble symmetries \cite{Dyson:1962es}:
\begin{itemize}
\item{Gaussian Orthogonal Ensemble (GOE): $\nu = 1, A_1 = 8/27$;}
\item{Gaussian Unitary Ensemble (GUE): $\nu = 2, A_2 = 4\pi$;}
\item{Gaussian Symplectic Ensemble (GSE): $\nu = 4, A_4 = 4\pi/\big(729\sqrt{3}\big)$.}
\end{itemize}
We present the probability distribution of the logarithmic ratio $\ln r_n$, which is given by
\bea
P(\ln r_n) = p(r_n)r_n,
\eea
see Fig. \ref{fig:level_statistic_N_compare}.
In the figure, we remove the subscript $n$ for $r_n$ for the convenience of reading.
As the system approaches chaotic behavior, the distribution tends to resemble the GOE distribution, consistent with our Hamiltonian preserving real symmetric properties.
The adjacent gap ratio is defined as
\bea
\tilde{r}_{n}=\frac{\min(r_{n-1},r_{n})}{\max(r_{n-1},r_{n})}.
\eea
After averaging the adjacent gap ratio, a Poisson distribution gives \cite{Atas:2013gvn}
\bea
\langle \tilde{r}\rangle \approx 0.386
\eea
whereas for GOE \cite{Atas:2013gvn}
\bea
\langle \tilde{r}\rangle \approx 0.536.
\eea
\\

\noindent
As shown in Fig. \ref{fig:level_statistic_N_compare}, the phase diagram of $\langle \tilde{r}_n\rangle$ indicates the transition between Poisson and GOE statistics for different values of the variance ratio $\langle U^2 \rangle / \langle t_h^2 \rangle$ in the 6-, 8-, and 10-orbital systems.
When this ratio approaches zero, the system becomes integrable.
However, for $\langle U^{2}\rangle/ \langle t_{h}^{2}\rangle>0.33$ in the 10-orbital case, the system exhibits GOE statistics, aligns with RMT predictions.
Moreover, we observe that the number of orbitals increases, the integrable regime tends to vanish, and the value of $\langle \tilde{r}\rangle$ approaches the GOE limit upon introducing the four-point interaction term.
In the thermodynamic limit, the system initially enters the integrable regime when $\langle U^2 \rangle$ vanishes first, but transitions into the chaotic regime if $\langle t_h^2 \rangle$ approaches zero first.
Thus, our key conclusion is that a non-zero variance of $U$ drives the system toward chaotic behavior.
This finding highlights the crucial role of interaction disorder in inducing quantum chaos within the orbital HK model.

\subsection{Spectral Form Factor}
\begin{center}
\begin{figure*}[ht]
\centering
\includegraphics[width=.9\linewidth]{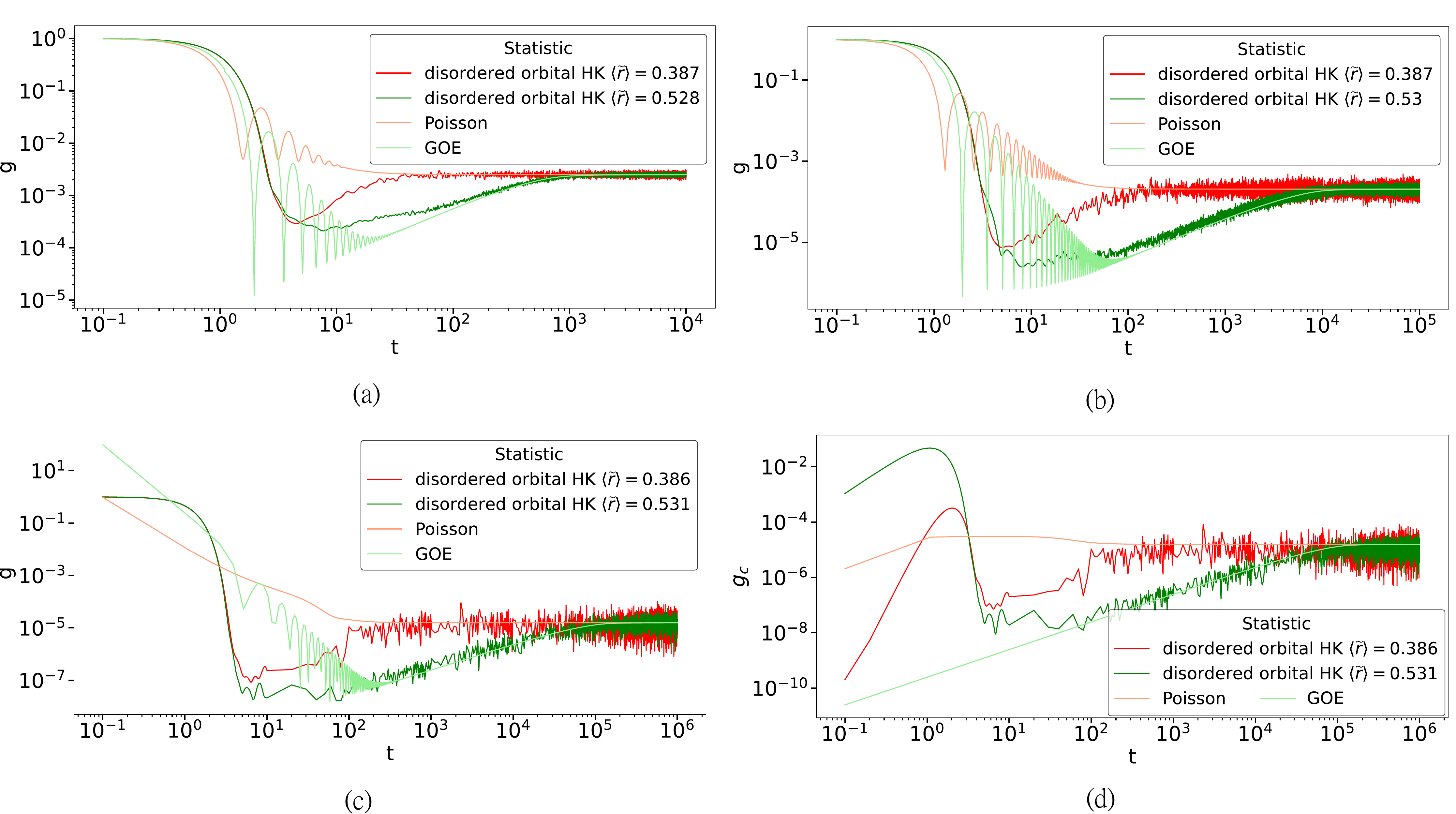}
\caption{SFF at infinite temperature for the disordered orbital HK model with 6, 8, and 10 orbitals at half-filling and zero total spin. The chaotic phase is characterized by $\langle \tilde{r} \rangle = 0.531$ with $\langle t_h^2 \rangle = 1$ and $\langle U^2 \rangle = 3.3$, and the integrable phase by $\langle \tilde{r} \rangle = 0.386$ with $\langle t_h^2 \rangle = 1$ and $\langle U^2 \rangle = 0$. (a) The results for the 6-orbital scenario are averaged across 500 random samples.
(b) The 8-orbital scenario, averaged across 50 random samples.
(c) The 10-orbital scenario, averaged across 5 random samples.
(d) The connected portion of the 10-orbital scenario.}
\label{fig:SFF_N6810_compare}
\end{figure*}
\end{center}

\noindent
While the level spacing statistic provides valuable GOE, the SFF helps offer supporting information.
The ramp feature observed in the SFF is attributed to long-range level repulsion \cite{Cotler:2016fpe,Dyer:2016pou}.
When balanced against the effects that keep the energy finite, this repulsion gives rise to a less rigid eigenvalue structure.
Consequently, due to its low rigidity, the SFF exhibits a ramp-shaped increase that linearly grows over time before transitioning into a plateau.
Note that $g(t)$ is dominated by the disconnected SFF $g_{d}(t)$ before the dip time and by the connected SFF $g_{c}(t)$ after the dip time.
In Ref. \cite{Cotler:2016fpe}, a dip-ramp-plateau behavior was observed for the SYK model, particularly in the ramp and plateau regions, with a subtle difference in early-time behavior.
For the disordered orbital HK model, the unfolded SFF patterns in integrable and chaotic cases at infinite temperature fall between the Poisson and GOE predictions of RMT, as shown in Fig. \ref{fig:SFF_N6810_compare}.
In the case where
\bea
\langle \tilde{r}\rangle=0.531,
\eea
the SFF exhibits the characteristic dip-ramp-plateau structure.
Furthermore, as seen in Fig. \ref{fig:SFF_N6810_compare}, with large orbitals considered, the SFF demonstrates the linear ramp as the theoretical prediction.

\section{OTOCs}
\label{sec:4}
\begin{center}
\begin{figure*}[ht]
\centering
\includegraphics[width=.95 \textwidth]{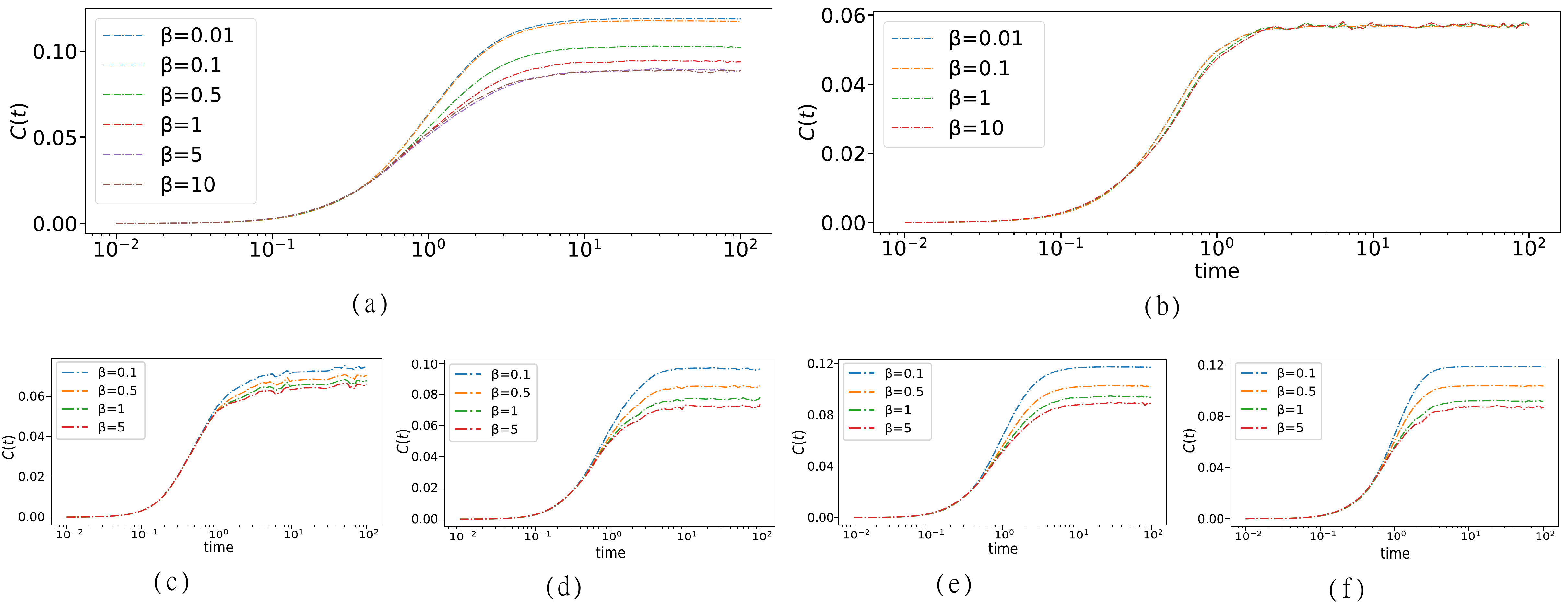}
\caption{Results for 30 random samples with 6 orbitals.
The OTOCs versus time on a log scale at different temperatures.
(a) GOE-level statistics with $\langle t_{h}^{2}\rangle = 1$ and $\langle U^{2}\rangle = 3.3$.
(b) Poisson-level statistics with $\langle t_{h}^{2}\rangle = 1$ and $\langle U^{2}\rangle = 0$.
For the Poisson case, the OTOCs calculations are totally temperature independent. For the GOE case, the late-time values of OTOCs are temperature dependent. $C(t)$ calculation with different system sizes:
(c) 2 orbitals with results averaged over 1000 random samples;
(d) 4 orbitals with 100 random samples;
(e) 6 orbitals with 30 random samples;
(f) 8 orbitals with only 1 random sample.
The results show that the system size does not affect the temperature-dependent late-time behavior of OTOCs.}
\label{fig:Ct_compare}
\end{figure*}
\end{center}

\begin{center}
\begin{figure*}[ht]
\begin{subfigure}[b]{1.\linewidth}
\centering
\includegraphics[width=1. \textwidth]{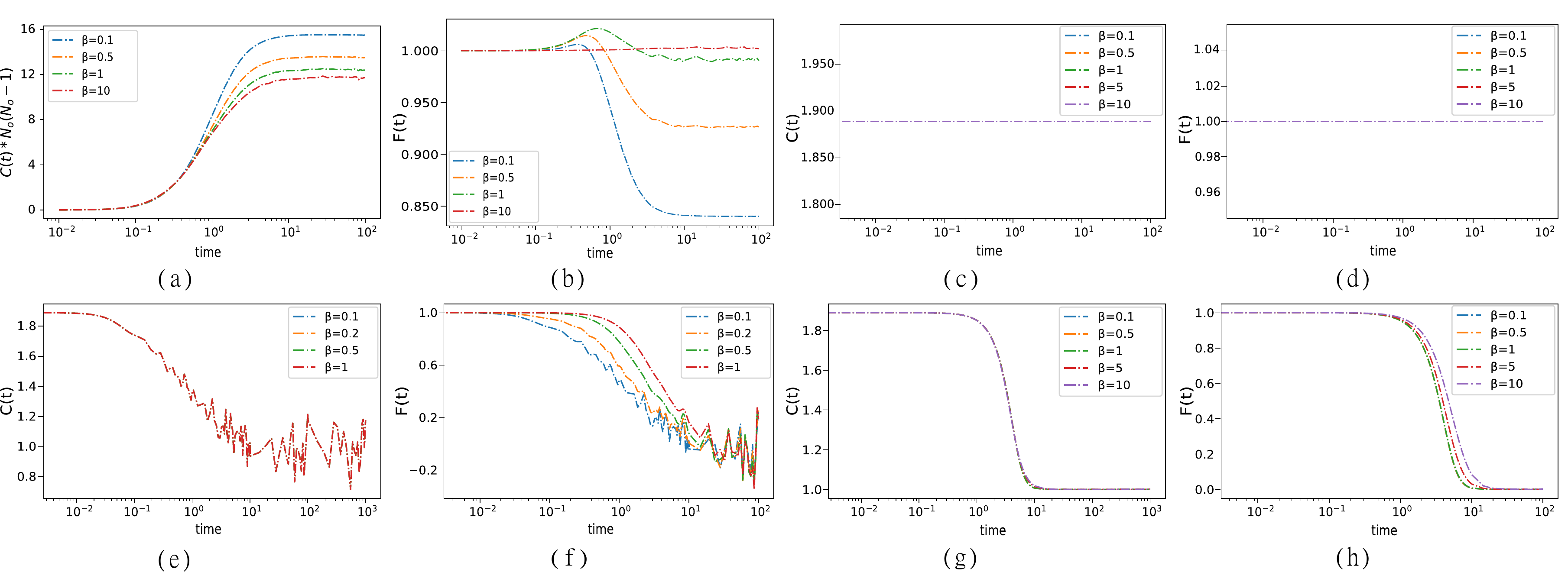}
\end{subfigure}
\caption{OTOCs calculations: $C(t)$ and normalized $F(t)$ for various models.
(a)(b) Disordered orbital HK model with $\langle t_{h}^2 \rangle = 1$ and $\langle U^2 \rangle = 3.3$, considering 6 orbitals over 30 random samples.
The number of number operators $N_o=12$.
(c)(d)   SYK$_2$ model with $\kappa = 1$ and $N = 18$ over 10 random samples.
(e)(f) Disorder-free SYK model with $N = 18$.
(g)(h) SYK$_4$ model with $N = 18$ over 10 random samples.}
\label{fig:Ct_Cregt_Ft_compare}
\end{figure*}
\end{center}

\noindent
To study the disordered orbital HK model, we choose operators as the number operators $n_{\alpha,\sigma}$ in the OTOCs
\begin{equation}
C(t) = \frac{1}{N_{o}(N_{o}-1)}\sum_{n_{\alpha,\sigma}\neq n_{\alpha',\sigma'}}-\langle[n_{\alpha,\sigma}(0),n_{\alpha',\sigma'}(t)]^{2}\rangle,
\end{equation}
where $N_{o}$ is the number of $n_{\alpha,\sigma}$.
We calculate $C(t)$ by averaging over random samples $\langle C(t)\rangle_{t_{h},U}$.
As shown in Fig. \ref{fig:Ct_compare}, in the Poisson distribution regime where $\langle t_{h}^{2}\rangle=1$ and $\langle U^{2}\rangle=0$, the OTOCs remain constant across different temperatures.
However, in the GOE distribution regime, with $\langle t_{h}^{2}\rangle=1$ and $\langle U^{2}\rangle=3.3$, the late time behavior of OTOCs increases with temperatures, as illustrated in Fig. \ref{fig:Ct_compare}. This behavior possibly arises from varying exponents during early times. Due to the numerical difficulties in calculating the Lyapunov exponent,
we only consider the late-time behavior of OTOC.
Additionally, as depicted in Fig. \ref{fig:Ct_compare}, the system size does not affect the temperature-dependent late-time behavior of OTOCs.
However, when we analyze the plateau values of OTOCs in many many-body models, such as the SYK$_{2}$ model, the disorder-free SYK model \cite{Lau:2020qnl}, and the SYK$_{4}$ model, we observe that the saturated values are temperature-independent across both integrable and maximally chaotic models, as illustrated in Fig. \ref{fig:Ct_Cregt_Ft_compare}.
Furthermore, we also examine the regularization issue through the regularized OTOCs,
\bea
F(t) = \mathrm{Tr}(yV(0)yW(t)yV(0)yW(t)),
\eea
where $y$ is defined by
\bea
y^{4} = \frac{1}{Z}e^{-\beta H},
\eea where
\bea
Z=\mathrm{Tr}(e^{-\beta H}),
\eea
as shown in Fig. \ref{fig:Ct_Cregt_Ft_compare}.
The normalized $F(t)$, i.e., $F(t)/F(0)$, shows that the saturated values tend to exhibit similar behavior in the SYK$_{2}$, disorder-free SYK, and SYK$_{4}$ models. This result aligns with the $C(t)$ calculation, suggesting that the relationship between the late-time behavior of OTOCs and temperature may not be a reliable indicator of chaotic properties in many-body systems.
However, a different behavior happens in the disordered orbital HK model.
Therefore, the late-time OTOCs, in general, depend on the regularization.

\section{Discussion and Conclusion}
\label{sec:5}
\noindent
In this paper, we investigated the adjacent gap ratio (a measure of energy level repulsion) \cite{Atas:2013gvn}, spectral form factor \cite{Brezin:1997rze}, and out-of-time-order correlators in the disordered orbital HK model.
The disordered orbital HK model, when Fourier transformed to position space, shows similarities to the SYK$_2$+SYK$_4$ model \cite{Garcia-Garcia:2017bkg}, a well-known system for studying quantum chaos and integrability. 
SYK$_2$ and disorder-free SYK models exhibit temperature independence in late-time OTOC behavior, similar to the SYK$_4$ model.
The disordered orbital HK model demonstrates dependence on regularization in its late-time behavior, suggesting that late-time OTOCs are not a reliable indicator of many-body chaos.
This implies that using OTOC saturation values alone to distinguish between integrability and chaos may not always work.
\\

\noindent
We delved into an intriguing area of quantum chaos, particularly regarding the transition between chaotic and integrable behavior within quantum systems.
Studying this transition through simpler models like the disordered orbital HK model is promising because it is less complex than models like the more involved SYK$_2$+SYK$_4$ model.
In classical mechanics, integrable systems can be solved exactly, with as many conserved quantities as degrees of freedom, leading to regular and predictable motions.
Chaotic systems, on the other hand, exhibit sensitive dependence on initial conditions and complex dynamics.
Extending this distinction to quantum systems introduces a challenge, as quantum chaos needs a straightforward analog to classical chaos.
\\

\noindent
The HK model is a lattice-based system where electrons hop between orbitals in a disordered manner \cite{Manning-Coe:2023etj,hatsugai1992exactly}, and can be modified with random variables to study NFL behavior and chaos. 
By tuning the disordered interaction strength in the disordered orbital HK model, we can induce a transition from integrable to chaotic behavior.
This transition is essential for understanding quantum chaos in more complex systems, particularly NFLs.
NFL behavior often arises in strongly correlated systems, where quasiparticle descriptions break down. 
The chaotic nature of some NFLs suggests a deeper connection between chaos and the breakdown of conventional many-body theory.
By studying the disordered orbital HK model, which might show NFL behavior under certain conditions, and comparing it with other chaotic models like SYK$_2$+SYK$_4$, we can explore how chaos influences the NFL state and its transitions.

\section*{Acknowledgments}
\noindent 
We would like to express our gratitude to Masaki Tezuka for his helpful discussion.
CTM would thank Nan-Peng Ma for his encouragement.
PYC acknowledges support from the National Science and Technology Council of Taiwan under Grant No. NSTC 113-2112-M-007-019.
Both PYC and YLL thank the National Center for Theoretical Sciences, Physics Division for their support.
YLL also thanks Chi-Ting Ho for insightful discussions. 
Computational resources were provided by Academia Sinica Grid Computing Centre, supported by Grant No. AS-CFII-112-103.


\end{document}